\documentclass[aps,pre,floatfix,twocolumn]{revtex4-1}
\usepackage{epsfig}
\usepackage{float}
\usepackage{verbatim}	 
\usepackage{amssymb}
\usepackage{bbold}
\usepackage{amsmath}
\usepackage{psfrag}
\usepackage{color}
\usepackage{ulem}
\newcommand{\bi}{\bibitem}

\usepackage[latin1]{inputenc}
\usepackage[english]{babel}
\usepackage[colorlinks, citecolor=blue, linkcolor=red, urlcolor=blue]{hyperref}
\usepackage{graphicx}
\usepackage{amscd}
\usepackage{eucal}
\usepackage{bm}
\usepackage{wrapfig}
\input {epsf}
\begin{document}

\title{Feasibility and Single Parameter Scaling of Extinctions in Large Ecological Communities}
\author{Philippe Jacquod}

\affiliation{Department of Quantum Matter Physics, University of Geneva, Geneva, Switzerland\\
Andlinger Center for Energy and the Environment, Princeton University, Princeton, USA \\
School of Engineering, University of Applied Sciences of Western Switzerland HES-SO, Sion, Switzerland}

\date{\today}

\begin{abstract}
Multispecies ecosystems modelled by generalized Lotka-Volterra equations exhibit stationary 
population abundances, where large number of species often coexist.
Understanding the precise conditions under which this is at all feasible and what triggers species extinctions
is a key, outstanding problem in theoretical ecology.  Using standard methods of random matrix theory, I 
show that distributions of species abundances are Gaussian at equilibrium, in the weakly interacting regime. 
One consequence is that feasibility is generically broken before stability, for large enough number of species. 
I further derive 
an analytical expression for the probability that $n=0,1,2,...$ species go extinct and conjecture that a single-parameter 
scaling law governs species extinctions. These results are corroborated by numerical simulations 
in a wide range of system parameters.
\end{abstract}

\maketitle

{\bf Introduction.} Many physical systems are so complex, that not only their state, but even their nature itself cannot be determined precisely. To deal with such systems, 
Wigner, Dyson and others constructed an extension of statistical mechanics, where either the Hamiltonian matrix, the time-evolution operator or, in the case of 
an open system, the scattering matrix belongs to an ensemble from which statistical properties can be evaluated~\cite{Wig58,Dys62}.
The resulting Random Matrix Theory (RMT)~\cite{RMT} has been very successful in describing the excited spectrum of nuclei~\cite{Bro81,Zel96}, of
small metallic nanograins~\cite{Gor65}, of quantum chaotic systems~\cite{Haake} and of strongly correlated, many-body systems~\cite{Mon93,Jac97,Kos18}, in calculating 
quantum corrections to electronic transport in mesoscopic systems~\cite{Bee97}, and in characterizing the structure of excited states in 
complex atoms~\cite{Fla94} and heavy nuclei~\cite{Bro81,Zel96} among others. 

A similar statistical approach has been advocated in theoretical ecology.
Multispecies ecosystems are standardly modeled as dynamical systems, whose degrees of freedom represent abundances for each of $S$ species~\cite{May07}. Time-evolution is governed by
coupled differential equations with at least $\propto S^2$ parameters -- e.g. the carrying capacity for each species, as well as pairwise interspecies interactions~\cite{May07} -- and even more if higher-order interactions are taken into account~\cite{Woo02,Wer03}. Determining all of them with appropriate accuracy is a hopeless task in large, biodiverse ecosystems.
But even if all parameters were known, computing analytically the evolution of species abundances, determining the existence, nature and stability of attractors 
and anticipating species extinctions can only be delegated to numerical studies already for moderate $S$~\cite{May07,Gar89,Fus00,Mcg08}.  RMT has proven to be an invaluable alternative in the study of
multispecies ecosystems, by 
assuming that interspecies interactions have a random distribution defined by its average and variance. 
Under this assumption, May conjectured that equilibrium states of large ecosystems are generically linearly unstable -- almost any small perturbation would push the dynamics away from them --  unless the variance of
interspecies interactions goes down $\propto S^{-1}$ or faster~\cite{May72}. 

May's work initiated a new, RMT approach to theoretical ecology~\cite{May72,Rie89,Jan03,All12,All15,Fyo16,Gri17,Fri17,Bir18,Bar18,Gal18,Sto18,Ros23,Cle23,Mam24,Jac25,Akj24}.
Focusing on the generalized Lotka-Volterra model [See Eq.~\eqref{lv} below],
RMT established in particular that, when interspecies interactions are weakly fluctuating, a single, globally stable equilibrium exists, while when fluctuations are sufficiently strong, 
the system enters a phase with multiple, coexisting equilibria~\cite{Rie89,Bir18,Ros23}.
How the transition happens between the two regimes depends on how species interact~\cite{All12,Fri17,Cle23,Ros23}. 
In particular, mutualistic or competitive interactions have been found to be generally destabilizing, while predator-prey interactions tend to stabilize ecosystems~\cite{All12,Cle23,Mam24}.
These, and other RMT results in theoretical ecology have been reviewed in Refs.~\cite{All15,Akj24}. 
May's work re-opened the debate on the common wisdom 
that complexity --  measured e.g. by biodiversity -- begets stability, an important, still open question~\cite{McA55,Nun80,Mcc00,Hat24,Agu24}. 

Stability is however only one side of the problem, and soon after May's work, it was pointed out that, when investigating 
species coexistence in mathematical models, stability and
feasibility --  the condition that all species abundances are positive -- must be considered on an equal footing~\cite{Rob74}.  
Coexistence from the combined point of view of stability and feasibility has been investigated within RMT in Refs.~\cite{Gri17,Ros01,Sto16,Dou18,Sto18,Biz21,Biz21,Akj22,Mar25,Cle23,Akj24}. 
A general, qualitative consensus seems to emerge, that feasibility is more restrictive than stability, though an analytical theory of feasibility is still lacking. 

In this paper, I use RMT to calculate average, variance, skewness and kurtosis of the distribution $P(N_i^*)$ of species abundances for equilibrium fixed-points 
of generalized Lotka-Volterra equations, in the regime of weak interaction and for $S \gg1$. I find that abundances follow a Gaussian distribution, with average $\overline{N}$ and variance $\Sigma_N$ 
determined by the average, the variance and the cross-diagonal correlator of interspecies interactions [See Eqs.~\eqref{randommatr} below]. This finding generalizes the analytical results of 
Ref.~\cite{Gem82} to nonvanishing values of $\gamma$ and $\mu$ in Eqs.~\eqref{randommatr}.
It furthermore  rigorously justifies later extensions of that work based on heuristic arguments, that abundances are normally distributed~\cite{Cle23,Akj24}.
From $P(N_i^*)$, I give an analytic expression for 
the probability that stationary solutions have $n=0,1,2,...$ species with negative abundances, as a function of the number $S$ of species. When $n \ge 1$, the fixed-point is unfeasible, and 
the key question is whether $P(N_i^*)$
has any connection with the extinction dynamics governed by the generalized Lotka-Volterra equations. 
I show analytically and confirm numerically that $P(N_i^*>0)$
gives the distribution of abundances for dynamically surviving species as long as the number $S_e$ of extinct species is small, $S_e \ll S$. 
In that regime, the analytic expression for the probability to have $n=0,1,2,...$ species with negative abundances gives the probability distribution for the number of extinct species.
I finally find numerically that the ratio of average and 
standard deviation of the Gaussian distribution of abundances gives a single parameter which governs a scaling law for the number of extinctions in generalized Lotka-Volterra equations,
in a surprisingly large set of parameters.

{\bf Model and Methodology.} I consider large Lotka-Volterra models defined by the following  $S$ coupled nonlinear ordinary differential equations~\cite{All15,Akj24}
\begin{equation}\label{lv}
 \dot N_i = N_i \Big( k_i- N_i - \sum_{j=1}^S {\mathbb A}_{ij} N_j \Big) \, , \; i=1, \ldots S \, .
\end{equation}
They generalize the Lotka-Volterra model~\cite{LV1,LV2} to mixed ecosystems with multiple species. 
They determine the time-evolution of the abundance $N_i(t)$ for each species labeled $i$, as a function of (i) its growth rate $k_i$ and (ii) interspecies interactions encoded 
in the elements of the real $S \times S$ matrix $\mathbb A$. Following RMT~\cite{RMT}, $\mathbb A$ belongs to an ensemble defined by the statistics 
of its matrix elements ${\mathbb A}_{ij}$, and I assume that the latter  
are normally distributed with  average and second moment given by
\begin{subequations}
\label{randommatr}
\begin{eqnarray} 
\langle {\mathbb A}_{ij}\rangle & = & \mu/S \;, \\
\label{xcorr}
\langle {\mathbb A}_{ij} {\mathbb A}_{kl} \rangle & = & \sigma^2 ( \delta_{ik} \delta_{jl}+\gamma \delta_{il} \delta_{jk})/S + (\mu/S)^2 \, .
\end{eqnarray}
\end{subequations}
Beside $\mu$ and $\sigma$, Eq.~\eqref{xcorr} introduces a third parameter $\gamma$ in our theory, whose meaning is the following. 
The interaction matrix element ${\mathbb A}_{ij} > 0 $ ($<0$) represents the loss (growth) rate of species $i$ due to its interaction with 
species $j$. Then, if ${\mathbb A}_{ij} > 0 $ ($<0$) and ${\mathbb A}_{ji} < 0$ ($>0$) the two species $i$ and $j$ are in a predator-prey 
relationship. If, on the other hand ${\mathbb A}_{ij} > 0 $ and ${\mathbb A}_{ji} > 0$ then the two species are in competition, while when
${\mathbb A}_{ij} < 0 $ and ${\mathbb A}_{ji} < 0$, they are in mutualistic symbiosis.  Thus $\gamma <0$ ($>0$) means that pairs 
with ${\mathbb A}_{ij}$ having the opposite (same) sign as ${\mathbb A}_{ji}$ dominate.
The cross-diagonal covariance parameter $\gamma \in [-1,1]$ tunes between exclusively predator-prey behaviors, $\gamma=-1$,
to exclusively competitive and/or mutualistic behavior, $\gamma=1$~\cite{caveat1}. 

Population abundances $N_i^*$ at the equilibrium solution of Eq.~\eqref{lv} are determined by $N_i^*=0$ for extinct species and
$k_i =  N_i^* + \sum_{j=1}^S {\mathbb A}_{ij} N_j^* $ for surviving ones. Given a set of parameters $\{ \vec{k}, \mathbb A\}$ 
it is hard to determine {\it a priori} which species will be extinct. A first procedure is to numerically time-evolve Eq.~\eqref{lv}
until an equilibrium is reached. Strictly speaking, extinctions are reached only asymptotically, $t \rightarrow \infty$. In practice, they can be 
defined consistently by introducing a low enough threshold and considering as extinct the species whose abundance go below that threshold~\cite{Jac25,Roy20}.   
A second procedure is to iteratively solve the set of linear equations
\begin{equation}\label{equi2}
{\vec N^*} = (\mathbb{1}+\mathbb{A})^{-1} \vec{k} \, ,
\end{equation}
by removing negative abundances at each iteration until a solution is reached with only positive abundances for $S_s \le S$ surviving species. It is generally not clear that the
two procedures lead to the same set of extinctions. 

Below I compare the first procedure with a simplified version of the second, where extinctions are conjectured to be limited to the set ${\cal S}_{\cal N}$ of species with negative
abundances obtained by solving Eq.~\eqref{equi2} for the full set of $S$ species. This presupposes that, removing ${\cal S}_{\cal N}$, i.e. solving Eq.~\eqref{equi2} over the corresponding lower dimensional space 
will provide a feasible solution. That a solution exists is likely -- the randomness of $\mathbb A$ ensures that $\mathbb{1}+\mathbb{A}$ is singular only at isolated points $(\sigma,\mu)$.
That it is feasible is however not guaranteed and needs to be checked.  I show analytically that the two procedures give the same distribution of abundances for surviving species and the same 
distribution for the number of extinct species to leading order in $S_e/S$. 

From the solution of Eq.~\eqref{equi2} in the full $S$-dimensional species space,
I calculate the four lower moments of species abundances over an ensemble of different interaction matrices $\mathbb{A}$ defined by Eqs.~\eqref{randommatr}.
The  approach is to expand the matrix inverse in Eq.~\eqref{equi2} in a Neumann series and calculate RMT ensemble averages term by term, keeping only the leading order
terms that survive when $S \rightarrow \infty$. I do this in two different ways. The first one expands Eq.~\eqref{equi2} directly in  the matrix ${\mathbb A}$. This expansion
converges if the spectral radius of $ \mathbb{A}$  satisfies $\rho_{\mathbb A}<1$. From Refs.~\cite{Som88,All12,Cle23}, one has $\rho_{\mathbb A} = {\rm max}[|\mu|, \sigma(1+|\gamma|)$.
The approach is therefore valid for $|\mu|<1$ and $\sigma(1+|\gamma|)<1$. 
The second one is to write ${\mathbb A}={\mathbb M} + \delta {\mathbb A}$, with the matrix average ${\mathbb M}_{ij}=\langle{\mathbb A}_{ij}\rangle =\mu/S$, and to
expand Eq.~\eqref{equi2} in  the matrix $\delta {\mathbb A}$. The two methods coincide for $\mu=0$, and the second one is advantageous when $\mu \gg1$, because its radius of convergence
is $\sigma(1+|\gamma|)<1$. In both cases, convergence of the expansion guarantees that one is in the regime
$\sigma < (1+\gamma)^{-1}$, of linear stability~\cite{May72,All12}. Using a Neumann expansion to extract species abundances from Eq.~\eqref{equi2} has been used in Ref.~\cite{Pet20} in conjunction with
an assumption of statistical independence. The result is a geometric resummation of the Neumann expansion, which is impossible from Eqs.~\eqref{avgfins} and \eqref{sigmu} unless $\gamma, \mu, \chi=0$.

Calculational details for both expansions are given in the Supplementary Information.
I use the first expansion to calculate the first four moments of $P(N^*_i)$  up to order ${\cal O}(\mu^{p_1} \sigma^{p_2})$, $p_1+p_2 \le 6$. How it
is applied in practice is illustrated  in the Supplementary Information. 
The second expansion  retains its validity for arbitrarily large values of 
$\mu$, however, it is  mathematically more intricate and calculations are limited to ${\cal O}(\sigma^4)$. The two approaches give 
compatible results where their respective regimes of validity overlap. 

{\bf Abundance Statistics.}
I assume that carrying capacities are distributed with average and variance given by $ \langle k_i \rangle = \kappa$ and  $\langle k_i k_j \rangle  - \langle k_i \rangle^2 = \chi^2 \delta_{ij}$,
i.e. $\chi=0$ corresponds to 
equal carrying capacities for all species. 
I obtain the distribution average
\onecolumngrid
\begin{eqnarray}\label{avgfins}
\overline N = \kappa [(1+\mu)^{-1} +(1+\mu)^{-2}\gamma \sigma^2+\left((1+\mu)^{-2}+(1+\mu)^{-3} \right) \gamma^2 \sigma^4  + 5 \gamma^3 \sigma^6  ]+ {\cal O}(\sigma^8,S^{-1})\, .
\end{eqnarray}
It is important to realize that the $\sigma^6$-term is the zero$^{\rm th}$ order term in an expansion in powers of $\mu$. All other terms, are obtained using either expansion, and resumming powers of $\mu$. 
Therefore, Eq.~\eqref{avgfins} is valid for $|\mu|<1$, if the $\sigma^6$-term is taken into account, or for arbitrary values of $\mu$ 
if the $\sigma^6$-term is neglected (in which case its accuracy vs. $\sigma$ is reduced). 

With the first Neumann expansion I further obtain the variance of abundances  
\begin{eqnarray}\label{sigmu}
\Sigma_N&=& \kappa^2 \, [(1-2 \mu+3 \mu^2-4 \mu^3 + 5 \mu^4) \sigma^2 + (1-2 \mu+3 \mu^2) \sigma^4 + (4-10 \mu + 18 \mu^2) \gamma \sigma^4 +(1+6 \gamma+14 \gamma^2) \sigma^6] \nonumber \\
&&  + 
 \chi^2 \, [1 + (1+2 \gamma) \sigma^2 + (1+4 \gamma+5 \gamma^2) \sigma^4 + (1+6 \gamma+ 14 \gamma^2 + 14 \gamma^3) \sigma^6]  + {\cal O}(\sigma^8,\sigma^6 \mu^2, ...S^{-1}) \, .
\end{eqnarray}
\twocolumngrid
\noindent It looks like the first three contributions in the term $\propto \kappa^2$ in Eq.~\eqref{sigmu} could be resummed as $(\sigma^2+\sigma^4)/(1+\mu)^2 +2 \gamma \sigma^4[(1+\mu)^{-2}+(1+\mu)^{-3}]$.
However, extending the second Neumann expansion to the variance is rather intricate and I have been unable to confirm this analytically. The fact that contributions to $\overline N$ and $\Sigma_N$ remain finite
for $S \rightarrow \infty$ directly follows from the choice of $S$-dependance in Eq.~\eqref{randommatr}. If desired, further $S$-dependance can be attributed to $\sigma$ and $\mu$, e.g. reflecting increased or reduced
interactions as the number of species increases. 

\begin{figure}[h]
\vspace*{-0.9cm}
\center
\includegraphics[width=1.02 \columnwidth]{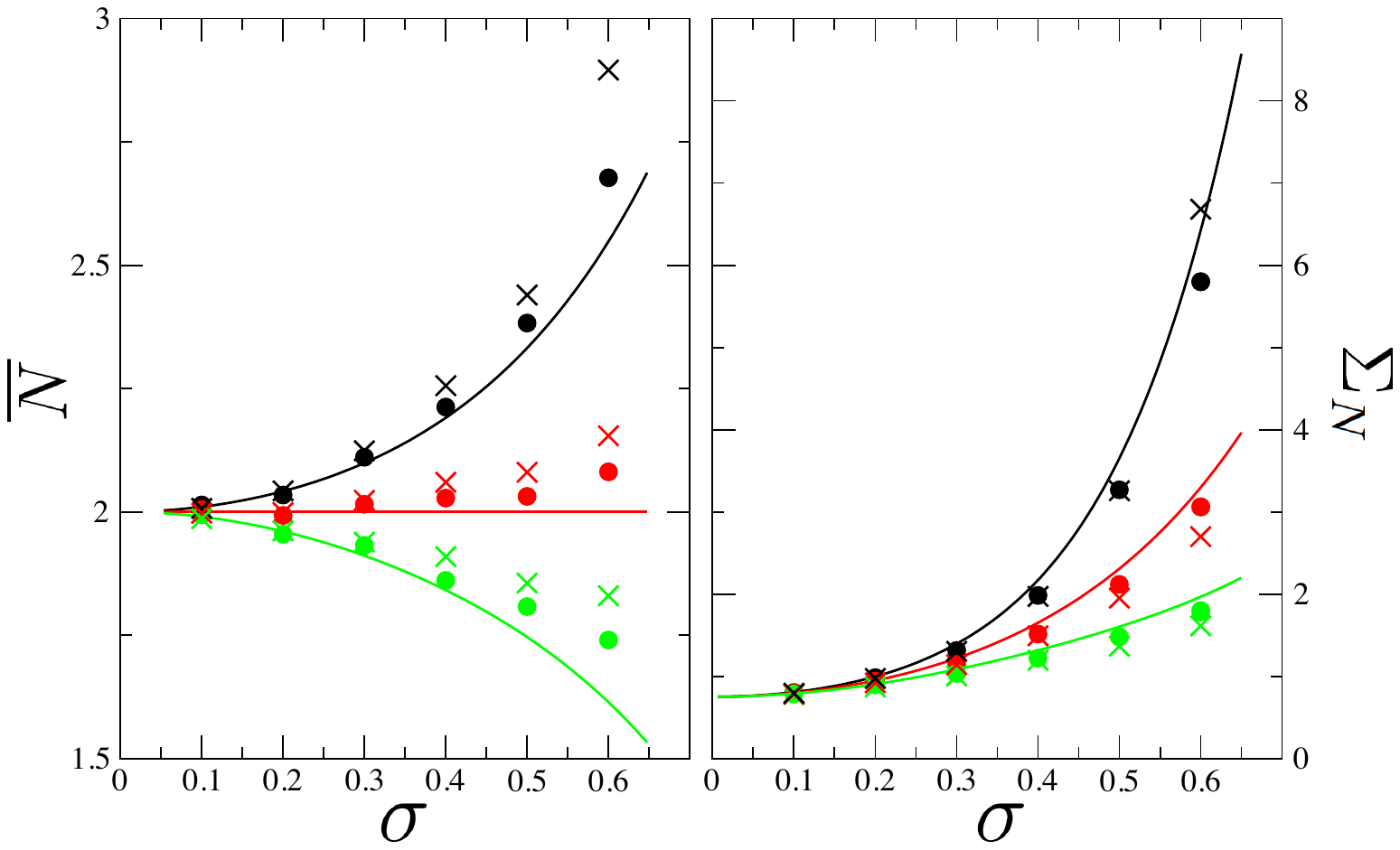}
\vspace{-13mm}
\caption{Average (left panel) and variance (right) of the abundance distribution at stable equilibria for Eq.~\eqref{lv}, 
$\mu=0$, and a random, homogeneous distribution $k_i \in [0.5,3.5]$, i.e. with $\kappa=2$ and $\chi=0.75$,
as a function of $\sigma$ and for $\gamma=-0.5$ (green), 0 (red) and 0.5 (black). 
Solid lines give the RMT predictions of Eqs.~\eqref{avgfins} and \eqref{sigmu}. Data correspond to averages over all species when the equilibrium of 
Eq.~\eqref{lv} has been reached for 500 realizations of the random interaction matrix $\mathbb A$ with $S=37$ ($\times$) and 157 (circles).\\[-6mm]
\label{fig:fig1}  }
\end{figure}
{\bf Results and Discussion.}
Fig.~\ref{fig:fig1} compares Eqs.~\eqref{avgfins} and  \eqref{sigmu} to numerically obtained results for the time-evolution of Eq.~\eqref{lv} to equilibrium. The agreement is very good for small to moderate values
of $\sigma$, but breaks down at larger values
of $\sigma \gtrsim 0.3$. For $\gamma >0$, this is so because one reaches the radius of convergence, $\sigma_c \simeq (1+\gamma)^{-1}$, of the Neumann expansion. For $\gamma<0$, terms $\propto \sigma^8$ and higher,
not included in our truncated perturbative expansion, become important. Additionally, the agreement becomes better 
at larger $S$, because corrections ${\cal O}(S^{-1})$ are neglected.

\begin{figure}[h!]
\vspace*{-0.8cm}
\center
\includegraphics[width=1.2 \columnwidth]{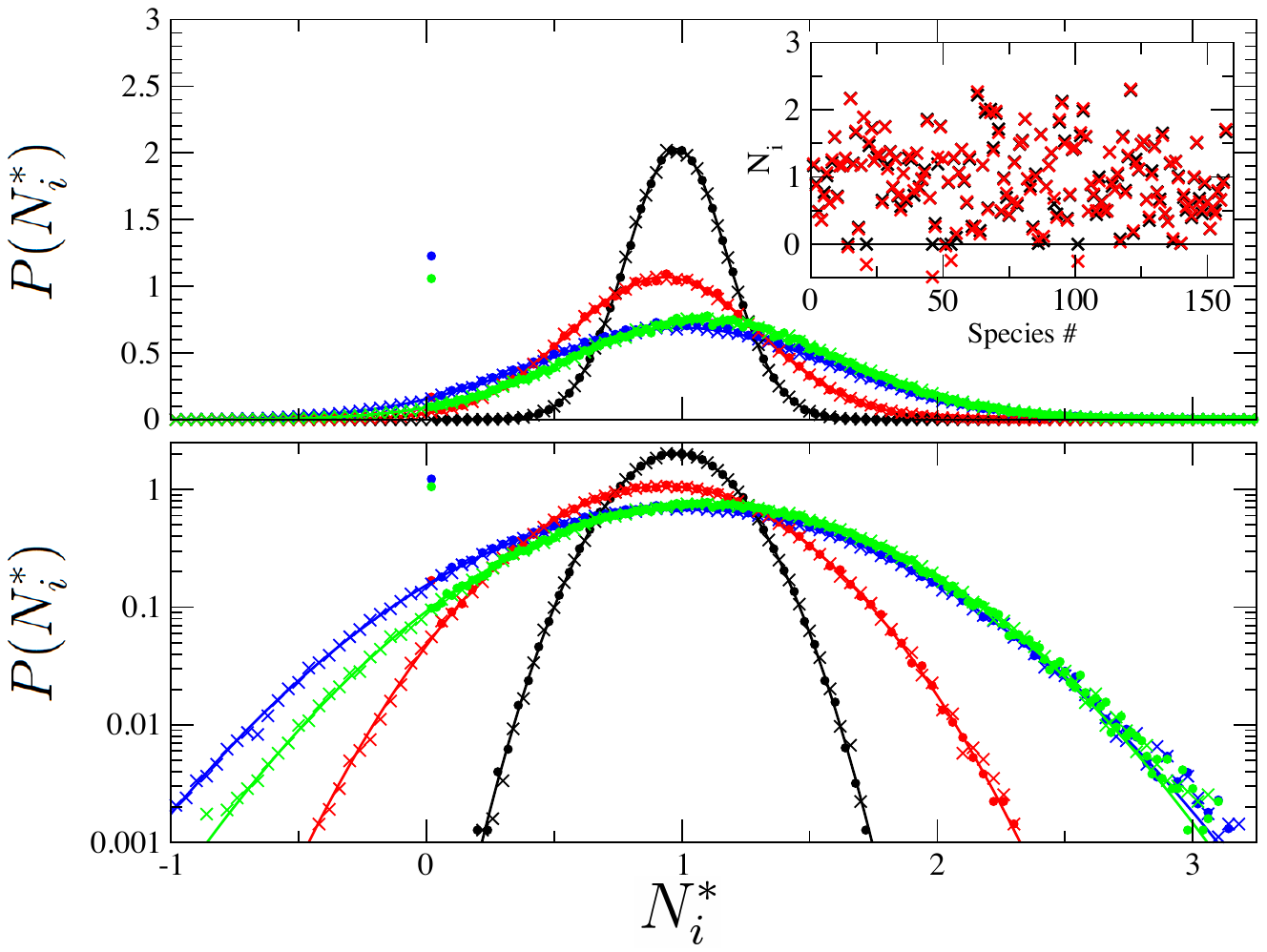}
\vspace{-11mm}
\caption{\label{fig:fig2} 
Normal (top) and semilog (bottom) plots of species abundance distributions for $\mu=0$, $k_i \equiv1$, and
$\sigma=0.2$, $\gamma=-0.5$  (black circles), $\sigma=0.4$, $\gamma=-0.5$ (red), $\sigma=0.5$, $\gamma=0$ (blue), $\sigma=0.4$, $\gamma=0.5$ (green).
Solid lines give Gaussian distributions with average ($\overline N = 0.98$, 0.93, 1.01 and 1.1) and variances ($\Sigma_N=0.038$, 0.138, 0.31, 0.29) 
given by Eqs.~\eqref{avgfins} and \eqref{sigmu}. Distributions are calculated from fixed-point solutions to Eq.~\eqref{lv} (dots) and from solutions
to Eq.~\eqref{equi2} ($\times$), over 
1000 realizations of the interaction matrix. The dots at $N_i^*=0$ correspond to $\simeq 6.7$ (blue), 3.3 (green) and 0.8 (red) extinctions per realization, in agreement 
with Eq.~\eqref{probNe}. 
Inset: Abundances $N_i$ at a fixed-point solution to Eq.~\eqref{lv} (black) and for Eq.~\eqref{equi2} (red) for the same realization of $\mathbb A$,
for $\mu=0$, $k_i \equiv1$, $\sigma=0.5$ and $\gamma=0$. There are six species with negative abundances at the fixed-point and six species with abundances 
below 10$^{-20}$ and still going down after 2 $\times 10^{7}$ Runge-Kutta iterations of Eq.~\eqref{equi2}.\vspace{-2mm}}
\end{figure}

In the Supplementary Information I show that the skewness vanishes and the kurtosis ${\cal K}=\langle (N_i^*-\overline N)^4\rangle/\Sigma_N^{2}=3 +{\cal O}(\sigma^8,S^{-1})$. 
Furthermore, standardized moments of order $2p>4$ have contributions already existing in lower moments, with combinatorial factors of $(2 p -1) !!$, relating the $2p^{\rm th}$ moment to the variance.
These results point toward a Gaussian 
distribution of abundances~\cite{Durrett}, which is confirmed in Fig.~\ref{fig:fig2}. Distributions obtained numerically from the time-evolution of Eq.~\eqref{lv} to equilibrium are equal to those obtained from 
Eq.~\eqref{equi2} for $N_i^* >0$. Both are furthermore Gaussian, with average and variance given by Eq.~\eqref{avgfins} and \eqref{sigmu}. This is true 
even when several species have negative abundances according to Eq.~\eqref{equi2}, in which case, they correspond to a peak at $N_i^*=0$ in the distribution for the equilibrium of Eq.~\eqref{lv}.
(See the blue, green and red dots at $N_i^*=0$ in Fig.~\ref{fig:fig2}, corresponding respectively to 6.5, 5.6 and 0.8 extinctions per realization). 
Thus there is a one-to-one correspondence between the extinct species obtained by letting Eq.~\eqref{lv} relax to equilibrium, and the species with negative abundances 
obtained from Eq.~\eqref{equi2}. This correspondence is further illustrated in the inset of Fig.~\ref{fig:fig2} for an individual realization of $\mathbb A$ with six extinctions.

The numerically observed correspondence between negative abundances at unfeasible fixed-points and extinctions can be analytically established. 
Consider as initial condition for Eq.~\eqref{lv} the unfeasible fixed-point, where species with negative abundances are removed. Surviving species will then evolve dynamically 
towards a new fixed-point, and  in the Supplementary Information I show that the shift in surviving abundances between the old and the new fixed-point is small, $\propto S_e/S$,
as long as the number of extinct species is small compared to the total number of species.

\begin{figure}
\vspace{-0.64cm}
\hspace*{-7mm}
\includegraphics[width=1.25 \columnwidth]{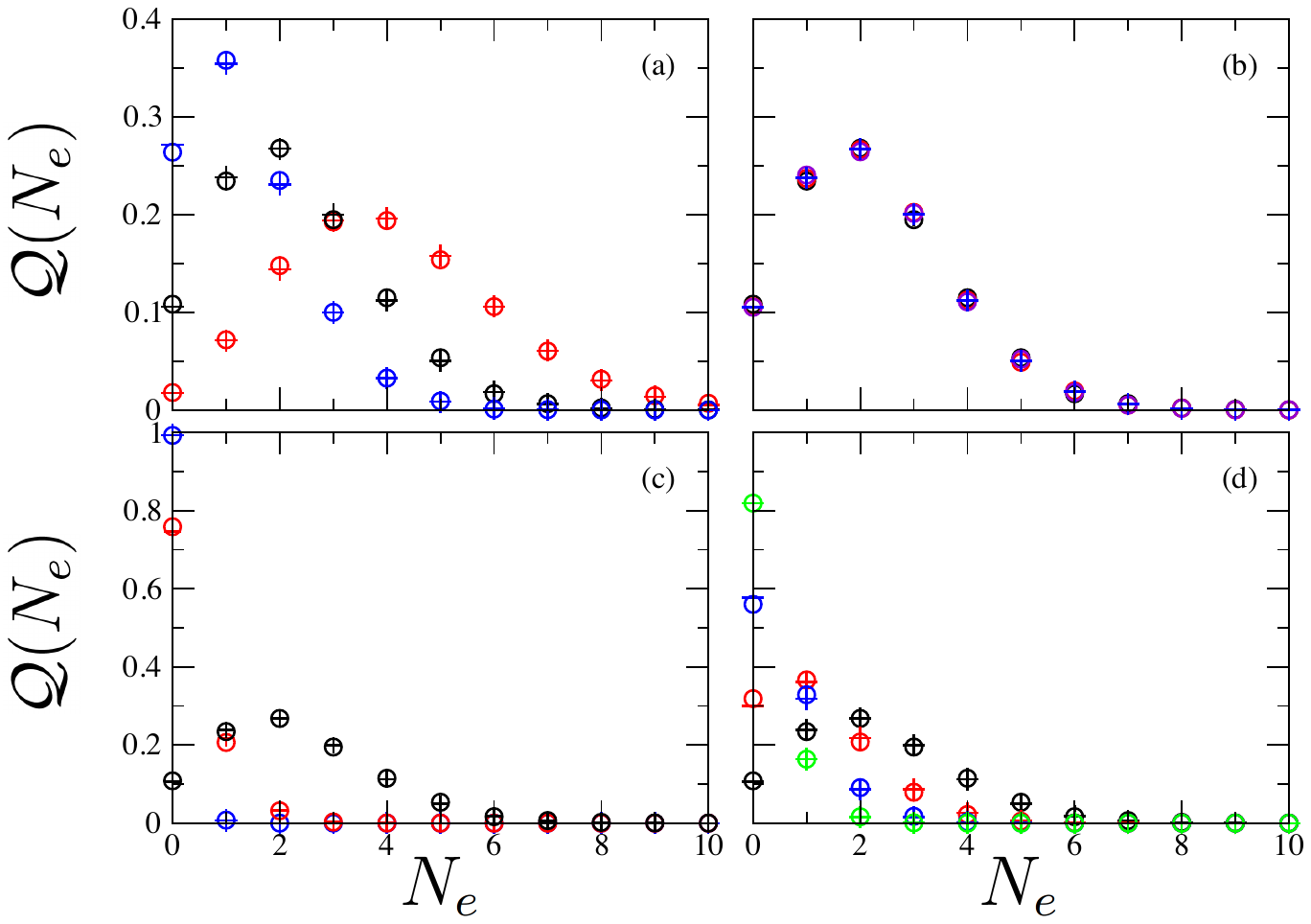}
\vspace{-12mm}
\caption{\label{fig:fig3} 
Distribution of number $N_e$ of extinct species, from the species with negative abundancies in Eq.~\eqref{equi2} (+) and from Eq.~\eqref{probNe} (dots), with $k_i \equiv1$. 
(a) $S=607$, $\mu=0$, $\sigma=0.35$ and $\gamma=-0.5$ (blue), 0 (black) and 0.5 (red); (b) $S=607$, $\sigma=0.35$, $\gamma=0$ and $\mu=0$ (black), 2 (red) and 4 (blue);
(c) $S=607$, $\mu=0$, $\gamma=0$ and $\sigma=0.23$ (blue), 0.29 (red) and 0.35 (black); (d) $\mu=0$, $\gamma=0$, $\sigma=0.35$ and $S=57$ (green), 157 (blue), 307 (red) and
607 (black). All distributions are calculated over 1000 different realizations of the interaction matrix $\mathbb A$. \\[-8mm]
}
\end{figure}

Species abundances at fixed-point equilibriums are thus normally distributed as $P(N_i^*)=\exp[-(N_i^*-\overline N)^2/2 \Sigma_N]/\sqrt{2 \pi \Sigma_N}$. The probability ${\cal Q}(N_e) $ to have $N_e$ species with negative abundances in
an ecosystem of $S$ species is therefore (see Supplementary Information)
\begin{subequations}
\label{probNe}
\begin{eqnarray}\label{prob1}
{\cal Q}(N_e) = 
\left(\begin{array}{c}
S \\ N_e
\end{array} \right) \,
F_-^{N_e} \, F_+^{S-N_e} \, , \\
F_{\pm}=\left(1 \pm {\rm erf}\left[\overline N/\sqrt{2 \Sigma_N}\right]\right)/2 \, ,
\end{eqnarray}
\end{subequations}
with the error function erf[x]. Fig.~\ref{fig:fig3} illustrates that ${\cal Q}(N_e)$ matches the distribution of species with negative abundances obtained from Eq.~\eqref{equi2}.
Furthermore, it shows that more negative abundances occur as $\sigma$ and $\gamma$ increase, but that ${\cal Q}(N_e)$ is insensitive to $\mu$. This is in agreement with Eqs.~\eqref{avgfins} and \eqref{sigmu} which 
give $\overline N^2/\Sigma_N$ independent of $\mu$ to leading order. More importantly, 
increasing $S$, every other parameter being equal, systematically leads to more negative abundances [Fig.~\ref{fig:fig3}{\color{red}d}]. This is so, because populating the left, negative tail
of the Gaussian distribution becomes more likely when $S$ increases [see the binomial coefficient in Eq.~\eqref{prob1}].  

\begin{figure}
\vspace{-0.5cm}
\hspace*{-0.7cm}
\includegraphics[width=1.2 \columnwidth]{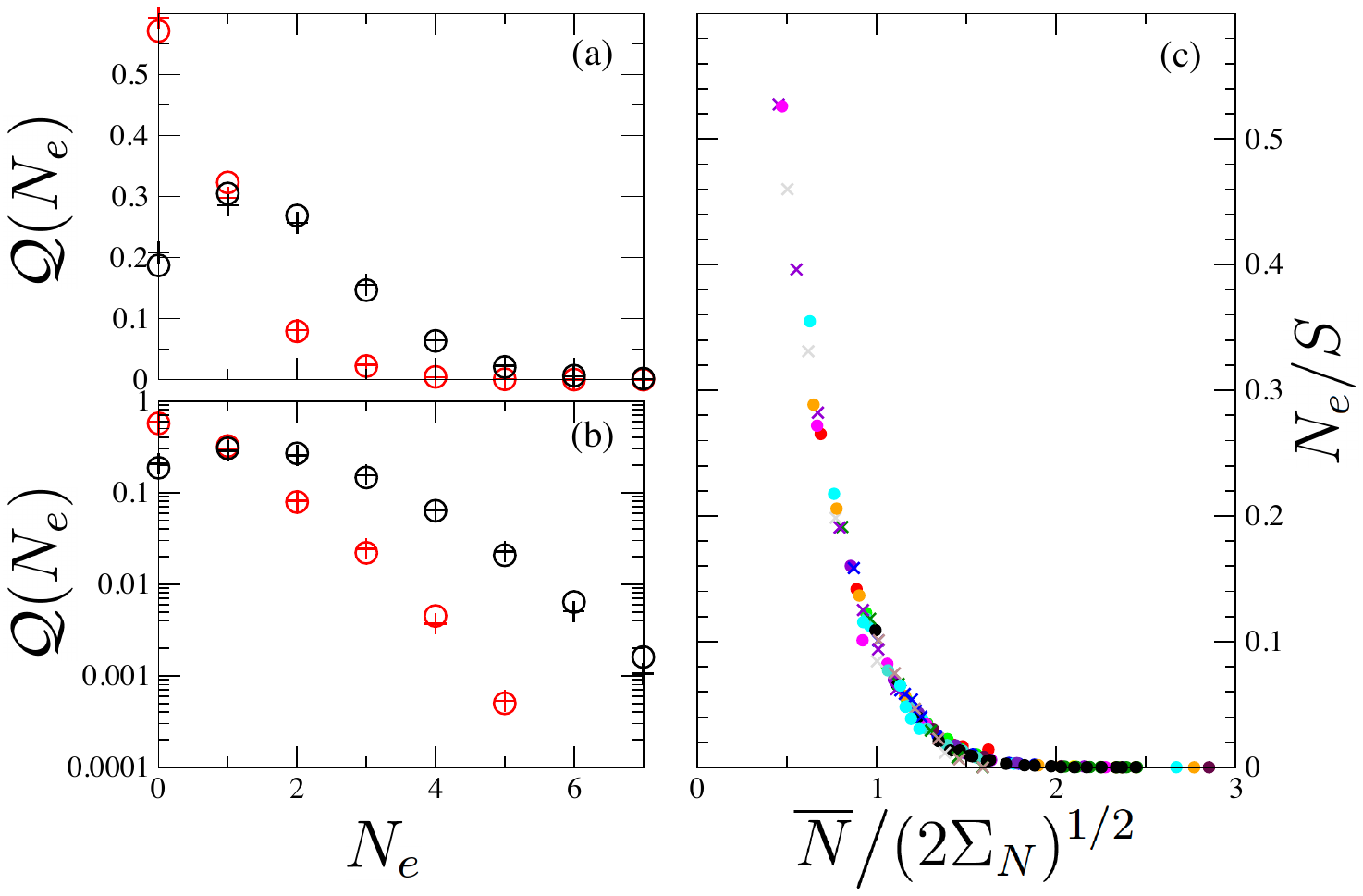}
\vspace{-12mm}
\caption{\label{fig:fig4} 
Normal (a) and semilog (b) plots of the distribution of number $N_e$ of extinct species, for $S=157$, $\mu=0$, $k_i \equiv1$, $\gamma=0$, $\sigma=0.35$ (red symbols) and 0.4 (black).
Distributions are calculated from fixed-point solutions obtained by time-evolving Eq.~\eqref{lv} (circles) and from solutions
to Eq.~\eqref{equi2} (+), over 1000 different realizations of the interaction matrix $\mathbb A$. 
(c) Scaling of the ratio of the average number of extinct species vs. ratio of the average over standard deviation of the 
abundance distribution, for 24 different sets of parameters, $\sigma \in [0,0.6]$, $\mu \in [0,1]$, $\gamma \in [-1,1]$, $S \in [100,300]$, and $k_i \equiv 1$ (circles)
as well as $k_i \in [0.5,3.5]$ (crosses). Averages are calculated over 100 to 500 realizations of the random matrix $\mathbb A$ for each set of parameter, by time-evolving Eq.~\eqref{lv} until a 
stationary solution is reached. Symbols of the same color correspond to sets varying either $\sigma$ or $\gamma$, all other parameters being fixed.
When time-evolving Eq.~\eqref{lv}, extinctions are defined at a threshold $N_c=10^{-20}$.\\[-8mm]
}
\end{figure}

Fig.~\ref{fig:fig2} suggested that species with negative abundances from Eq.~\eqref{equi2} are those that go extinct under the dynamics of Eq.~\eqref{lv}. This is corroborated in the two left panels of 
Fig.~\ref{fig:fig4}, which compare ${\cal Q}(N_e)$ of Eqs.~\eqref{probNe} with the distribution of extinctions obtained from time-evolving Eq.~\eqref{lv} -- the two distributions
are the same, even when there are several extinctions on average and a $\simeq 15 \%$ probability to have 3 or more extinctions. The two distributions remain the same at larger $\sigma$ and $\gamma$ (not shown).
Note that values for $\overline N$ and $\Sigma_N$ need to be adapted from Eqs.~\eqref{avgfins} and \eqref{sigmu} when $\sigma \gtrsim 0.3-0.4$. 
These results show that species extinctions can be statistically predicted by Eqs.~\eqref{probNe}, together with Eqs.~\eqref{avgfins} and \eqref{sigmu}. 

An important consequence of Eqs.~\eqref{probNe} is that species extinctions depend only on $\overline N/\sqrt{2 \Sigma_N}$. 
The right panel of Fig.~\ref{fig:fig4} demonstrates the existence of a single-parameter scaling law governing the distribution of extinctions -- the ratio $N_e/S$ falls on a single curve for widely varying 
system parameters, when plotted as a function of the scaling parameter $\eta = \overline N/\sqrt{2 \Sigma_N}$. This scaling is quite remarkable, as it applies even for $N_e/S \gtrsim 0.5$,
when more than half of the species have gone extinct.

{\bf Conclusion.} I have shown that RMT correctly predicts the distribution of species abundances for equilibrium solutions of multispecies Lotka-Volterra ecosystems. 
The theory predicts the emergence of unfeasibility and the distribution of species with negative abundances in solutions to Eq.~\eqref{equi2}.
Its range of validity lies inside the stability boundary of fixed-point solutions, and therefore demonstrates that feasibility generically breaks down before stability. Because the 
theory is statistical, it cannot rule out that the opposite occurs in rare, specific instances, however I have never  observed this numerically. 

I further made a one-to-one connection between species with negative abundances for Eq.~\eqref{equi2} and those that get extinct under the dynamics of Eq.~\eqref{lv}. 
This connection is rigorous as long as the relative number of extinct species is small, $S_e/S \ll 1$. This is  shown in the Supplementary Information.
An important consequence of this connection and of the Gaussianity of the
abundances distribution is the numerically confirmed conjecture that extinctions obey a single-parameter scaling law. 

Compared to earlier works on feasibility~\cite{Gri17,Ros01,Sto16,Dou18,Sto18,Biz21,Biz21,Akj22,Cle23,Akj24,Mar25}, the present work goes one step further as conditions for feasibility are derived vs. all RMT parameters -- in particular the cross-diagonal correlator
$\gamma$ can take any value, while earlier works
considered specific choices $\gamma=0$ or $\gamma=1$ with occasional heuristic extensions  to other values~\cite{Mar25}. Additionally, this works considers nonconstant carrying capacities, beyond the standard assumption
$k_i \equiv 1$. It is clear from Eq.~\eqref{sigmu} that distributed $\{ k_i \}$ with a finite, nonzero variance $\chi^2$ significantly impact the conditions for feasibility. 
In this respect, Eqs.~\eqref{probNe} gives the probability to have no extinction as ${\cal Q}(0) = \left[\left(1 + {\rm erf}\left[\overline N/\sqrt{2 \Sigma_N}\right]\right)/2\right]^S$, which vanishes for fixed values of 
$\overline N/\sqrt{2 \Sigma_N}$ in the limit $S \rightarrow \infty$. For $k_i \equiv 1$, Refs.~\cite{Dou18,Biz21} showed that a transition occurs at $\sigma^*=(2 \log S)^{-1/2}$, where there are almost surely no extinction for $\sigma < \sigma^*$ 
and almost surely extinct species for $\sigma > \sigma^*$.
Eqs.~\eqref{avgfins} and \eqref{sigmu} suggest that this  remains the case if $\chi \sim (2 \log S)^{-1/2}$ so that the distribution of growth rates becomes sharper and sharper with $S$. When instead the distribution of $k_i$ keeps 
a finite width as $S$ grows, extinctions are inevitable.

The presented theory 
is perturbative in $\sigma$ and valid for $S >>1$, but makes no further assumption. I have found numerically that theoretical predictions
work well already for moderately large ecosystems $S \approx 30$.

{\bf Acknowledgment.}
I am thankful to Roland Willa for discussions on probabilities of extinctions.

\clearpage
\onecolumngrid
\begin{center}
\textbf{Feasibility and Single Parameter Scaling of Extinctions in Large Ecological Communities \\ Supplemental Information}
\end{center}
\setcounter{equation}{0}
\setcounter{figure}{0}
\setcounter{table}{0}
\setcounter{section}{0}
\makeatletter
\renewcommand{\theequation}{S\arabic{equation}}
\renewcommand{\thefigure}{S\arabic{figure}}
\renewcommand{\thesection}{S\arabic{section}}
\renewcommand{\bibnumfmt}[1]{[S#1]}

In this Supplementary Information section I briefly sketch the two Neumann expansions I use to solve 
\begin{equation}\label{equi2}
{\vec N^*} = (\mathbb{1}+\mathbb{A})^{-1} \vec{k} \, .
\end{equation}
The general method is the following. 
Inserting the Neumann series $(\mathbb{1}+\mathbb{A})^{-1} = \sum_p  (-1)^p \mathbb{A}^p$ into Eq.~\eqref{equi2} leads to 
\begin{equation}\label{expandN}
N^*_i = \sum_{p=0}^\infty \sum_{j=1}^S (-1)^p [\mathbb{A}^{p}]_{ij} k_j \, .
\end{equation}
Accordingly, the $m^{\rm th}$ moment $\langle (N^*_i)^m \rangle$ of the distribution $P(N^*_i)$ of abundances over the ensemble of random matrices $\mathbb{A}$ defined by Eqs.~({\color{red}2}) in the main text
reads
\begin{equation}\label{equin}
\langle (N^*_i)^m \rangle = \sum_{p_1, ... p_m} \sum_{j_1,... j_m} (-1)^{\sum_n p_n} \Big\langle \prod_{n=1}^m [\mathbb{A}^{p_n}]_{i j_n} k_{j_n} \Big\rangle \, .
\end{equation}
In this work, I assume that the distributions of $\mathbb{A}_{ij}$ and of $k_j$ are uncorrelated. Under this assumption,  I obtain
$\big\langle \prod_{n=1}^m [\mathbb{A}^{p_n}]_{i j_n} k_{j_n} \big\rangle = \big\langle  [\mathbb{A}^{p_1}]_{i j_1} \, [\mathbb{A}^{p_2}]_{i j_2}...  [\mathbb{A}^{p_m}]_{i j_m} \big\rangle \big\langle k_{j_1} k_{j_2} ... k_{j_m}\big\rangle$.
To calculate  averages over the interaction matrix ensemble, I use 
Eqs.~({\color{red}2}) and take the average over the carrying capacities as
\begin{eqnarray}
\label{kdistr}
\langle k_i \rangle = \kappa \, , \;\;\;\;\;\;\;\;\;\;\;
\langle k_i k_j \rangle  = \chi^2 \delta_{ij} + \kappa^2\, . 
\end{eqnarray}

\section{RMT calculation of the average  abundance} 

\subsection{$\mu=0$}

From Eq.~\eqref{equin},  the average species abundance is given by 
\begin{eqnarray}\label{avg0}
\overline N \equiv \langle N_i^* \rangle &=&  \sum_{p} \sum_{j} (-1)^p  \Big\langle [(\mathbb{A}^{p}]_{i j} k_{j} \Big\rangle 
= \sum_{p} \sum_{j} (-1)^p  \Big\langle [\mathbb{A}^{p}]_{i j} \Big\rangle \Big\langle k_{j} \Big\rangle
= \kappa  \sum_{p} \sum_{j} (-1)^p  \Big\langle [\mathbb{A}^{p}]_{i j} \Big\rangle \, ,
\end{eqnarray}
where the average is taken over both a RMT distribution of interaction matrices as well as a distribution of carrying capacities, under 
 the assumption that the two averages are uncorrelated. 

I first calculate contributions up to order ${\cal O}(\sigma^6)$ 
for $\mu=0$. They read
\begin{eqnarray}\label{avg1}
\overline N &=&  \kappa \Big(1+\sum_{l,j} \big\langle \mathbb{A}_{il} \mathbb{A}_{lj} \big\rangle + \sum_{l_1,l_2,l_3,j} \big\langle \mathbb{A}_{il_1} \mathbb{A}_{l_1 l_2} \mathbb{A}_{l_2 l_3} \mathbb{A}_{l_3j} \big\rangle
+ \sum_{l_1,l_2,l_3,l_4,l_5,j} \big\langle \mathbb{A}_{il_1} \mathbb{A}_{l_1 l_2} \mathbb{A}_{l_2 l_3} \mathbb{A}_{l_3 l_4} \mathbb{A}_{l_4 l_5} \mathbb{A}_{l_5j} \big\rangle \Big) + {\cal O}(\sigma^8) \, .
\end{eqnarray}
In particular, only terms with even powers of $\mathbb A$ contribute to the RMT average. The quadratic term is straightforwardly calculated using Eqs.~({\color{red}2}) in the main text,
\begin{eqnarray}\label{contractions2}
\sum_{l,j} \big\langle \mathbb{A}_{il} \mathbb{A}_{lj} \big\rangle & = &  (\sigma^2/S) \sum_{l,j} (\gamma \delta_{ij} + \delta_{il} \delta_{lj}) = \gamma \sigma^2 + {\cal O}(S^{-1}) \, ,
\end{eqnarray}
because only the first contribution has a sum surviving the Kronecker, which gives a factor $S$. To calculate the quartic and sextic contributions,
one needs to identify the pairwise contractions of indices implied by Eq.~({\color{red}2b}) in the main text, that give the dominant order in $S$. The rule is that $p^{\rm th}$ order contributions have $p$ sums and prefactors $\propto \sigma^p/S^{p/2}$. 
Each index contraction removes one sum and therefore only terms with at most $p/2$ independent index contractions give finite contributions in the limit $S \rightarrow \infty$. It turns out that $p/2$ is the minimal possible
number of index contractions. The quartic term reads
\begin{eqnarray}\label{contractions4}
\sum_{l_1,l_2,l_3,j} \big\langle \mathbb{A}_{il_1} \mathbb{A}_{l_1 l_2} \mathbb{A}_{l_2 l_3} \mathbb{A}_{l_3j} \big\rangle & = &
\sum_{l_1,l_2,l_3,j} \Big(\big\langle \mathbb{A}_{il_1} \mathbb{A}_{l_1 l_2}  \big\rangle \big\langle  \mathbb{A}_{l_2 l_3} \mathbb{A}_{l_3j} \big\rangle + \big\langle \mathbb{A}_{il_1} \mathbb{A}_{l_3 j}  \big\rangle \big\langle  \mathbb{A}_{l_1 l_2} \mathbb{A}_{l_2l_3} \big\rangle
+\big\langle \mathbb{A}_{il_1} \mathbb{A}_{l_2 l_3}  \big\rangle \big\langle  \mathbb{A}_{l_1 l_2} \mathbb{A}_{l_3j} \big\rangle \Big) \nonumber \\
&=& 
\gamma^2 (\sigma^4/S^2) \sum_{l_1,l_2,l_3,j} \big(\delta_{il_2} \delta_{l_2j} + \delta_{l_1l_3} \delta_{ij}\big) + {\cal O}(S^{-1}) = 2 \gamma^2 \sigma^4 + {\cal O}(S^{-1}) \, ,
\end{eqnarray}
\begin{wrapfigure}{r}{0.42\textwidth}
\hspace{-4mm}
\vspace{-6mm}\includegraphics[width=0.48 \columnwidth]{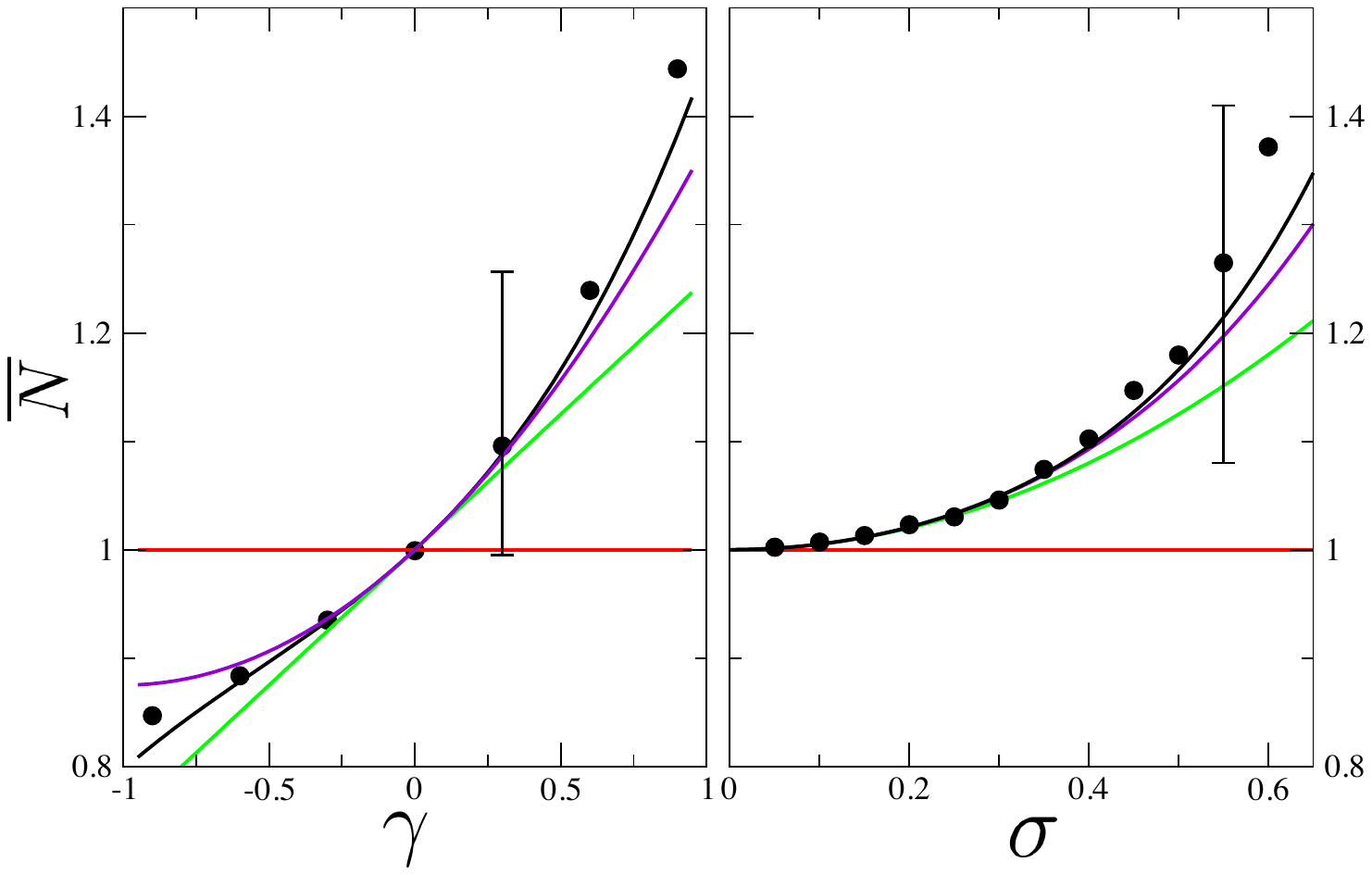}
\vspace{-9mm}
\caption{Convergence of the series expansion for $\mu=0$ and $k_i \equiv \kappa =  1$. The solid curves give Eq.~\eqref{avg1fin}  truncated at the zeroth (red), second (green), fourth (violet) and sixth (black) order in $\sigma$.
\label{fig:figS1}  }
\end{wrapfigure}
where the third term on the right hand side of the first line requires at least three index contractions, and therefore gives a subdominant, ${\cal O}(S^{-1})$ contribution.
Up to and including the quartic term, relevant contractions are easily identified. 
The procedure is similar for the sextic term, for which there are more contractions. I relied on a symbolic algebra code in parallel to the analytical calculation, 
to guarantee that all contributions are properly taken into account. 
For the sextic term in Eq.~\eqref{avg1}, there are five relevant contractions and  I obtain,
\begin{eqnarray}\label{contractions6}
\sum_{l_1,l_2,l_3,l_4,l_5,j} \big\langle \mathbb{A}_{il_1} \mathbb{A}_{l_1 l_2} \mathbb{A}_{l_2 l_3} \mathbb{A}_{l_3 l_4} \mathbb{A}_{l_4 l_5} \mathbb{A}_{l_5j} \big\rangle = 5 \gamma^3 \sigma^6 + {\cal O}(S^{-1})\, . \:\:
\end{eqnarray}
Putting Eqs.~(\ref{avg1}--\ref{contractions6}) together, I obtain
\begin{eqnarray}\label{avg1fin}
\overline N &=&  \kappa \Big(1+\gamma \sigma^2+2 \gamma^2 \sigma^4+5 \gamma^3 \sigma^6 \Big) + {\cal O}(\sigma^8,S^{-1}) \, .
\end{eqnarray}
Fig.~\ref{fig:figS1} illustrates the convergence of the series expansion, Eq.~\eqref{avg1fin}, to the numerically obtained results, as orders in $\sigma$ are added. \\[-10mm]

\subsection{$\mu \ne 0$}

I use two different approaches to extend this result to $\mu \ne 0$. Both are based on a Neumann expansion of Eq.~\eqref{equi2}
\begin{eqnarray}\label{neumann}
(\mathbb{X}+\mathbb{Y})^{-1} = (\mathbb{X})^{-1} \sum_{p=0}^\infty (-1)^p \left(\frac{\mathbb Y}{\mathbb X} \right)^p \, . \\[-10mm]
\nonumber
 \end{eqnarray}

\subsubsection{First expansion}\label{firstexpan}

The first Neumann expansion takes ${\mathbb X} = {\mathbb 1}$ and ${\mathbb Y}= {\mathbb A}={\mathbb M} + {\delta \mathbb A}$, with ${\mathbb M}_{ij}=\mu/S$,
and ${\delta \mathbb A}$ containing the zero-average fluctuations of interspecies interactions, i.e. with $\langle \delta \mathbb A_{ij} \rangle=0$ and  $\langle \delta \mathbb A_{ij} \, \delta \mathbb A_{kl}  \rangle= \sigma^2 ( \delta_{ik} \delta_{jl}+\gamma \delta_{il} \delta_{jk})/S$.
One has
\begin{eqnarray}\label{avgmu}
\overline N &=&  
\kappa  \sum_{p} \sum_{j} (-1)^p  \Big\langle [({\mathbb M}+\delta \mathbb{A})^{p}]_{i j} \Big\rangle \, ,
\end{eqnarray}
All terms with even powers of $\delta \mathbb A$ contribute to the RMT average, as before, and for $\mu \ne 0$, one additionally have terms with odd $p$ and odd powers of $\mathbb M$. That each average interaction comes with  $S$
in the denominator -- and not $S^{1/2}$ as the fluctuations of interaction -- still leaves finite contributions that survive the $S \rightarrow \infty$ limit, because they are not associated with Kronecker pairing of indices. 
To illustrate this I calculate the cubic contribution to $\overline N$. I obtain
\begin{eqnarray}\label{cubic}
(-1)^3 \sum_{l_1,l_2,j} \big\langle \mathbb{(M+\delta A)}_{il_1} \mathbb{(M+\delta A)}_{l_1 l_2} \mathbb{(M+\delta A)}_{l_2j} \big\rangle 
&=& -\mu^3 -(\mu/S) \sum_{l_1,l_2,j} \Big(\big\langle  \delta\mathbb{A}_{l_1 l_2}  \delta\mathbb{A}_{l_2j}\big\rangle + \big\langle \delta\mathbb{A}_{il_1} \delta\mathbb{A}_{l_1 l_2} \big\rangle + \big\langle  \delta\mathbb{A}_{i l_1}  \delta\mathbb{A}_{l_2j}\big\rangle  \Big)
\nonumber \\
&=& -\mu^3 -\gamma (\sigma^2/S) (\mu/S) \sum_{l_1,l_2,j} \Big( \delta_{l_1j}+\delta_{il_2} \Big) +{\cal O}(S^{-1}) \nonumber \\
&=& -\mu^3 - 2 \gamma \mu \sigma^2 +{\cal O}(S^{-1})  \, ,
\end{eqnarray}
where the first line is obtained by realizing that only even powers of $\delta {\mathbb A}$ survive RMT average. Applying this procedure to all terms up to and including $p=6$ in Eq.~\eqref{avgmu}, I obtain
\begin{eqnarray}\label{avgfin1}
\overline N = \kappa \,  \big[1-\mu+\mu^2-\mu^3+\mu^4-\mu^5+\mu^6 + \gamma \sigma^2 (  1-2 \mu+3 \mu^2- 4 \mu^3+ 5\mu^4 ) + \gamma^2 \sigma^4  (  2-5 \mu+9 \mu^2) + 5 \gamma^3 \sigma^6 \big] \, , 
\end{eqnarray}
where terms ${\cal O}(\mu^{p_1} \sigma^{p_2},S^{-1})$ with $p_1+p_2 > 6$ have been neglected. It is straightforward to see that 
when considering only the average interaction in all terms in Eq.~\eqref{avgmu}, one gets $\sum_{p+0}^\infty (-1)^p \mu^p = (1+\mu)^{-1}$. 
Still straightforward, terms $\propto \sigma^2$ that survive the $S \rightarrow \infty$ limit may acquire only a single Kronecker from the RMT average with Eqs.~({\color{red}2}) in the main text. Therefore, RMT contractions can be taken only over 
pairs of consecutive matrices  $\mathbb{M+\delta A} \rightarrow \delta \mathbb{A}$ in Eq.~\eqref{avgmu} that already share one index. There are $p-1$ of them, so that the $\sigma^2$ contribution gives
$\sum_p (-1)^p (p-1) \mu^{p-2} \sigma^2 = \sigma^2/(1+\mu)^2$. All $\sigma^4$ terms can finally be resummed in a similar, though a bit more intricate way: First, counting arguments similar to those after Eq.~\eqref{contractions2}
imply that only pairings with two Kroneckers in the RMT average with Eqs.~({\color{red}2}) in the main text
survive in the limit $S \rightarrow \infty$. Such contributions correspond to two independent pairs of consecutive $\mathbb M + \delta {\mathbb A}\rightarrow \delta \mathbb{A}$ in Eq.~\eqref{avgmu}. Furthermore, when these pairs are consecutive,
a multiplicative factor of 2 has to be taken into account, the origin of which is explained in Eq.~\eqref{contractions4}. Taking all this into account I end up with $p(p-3)/2$ contributions giving $\sum_p (-1)^p p(p-3)/2 \;  \mu^{p-4} \sigma^4 = \sigma^4[1/(1+\mu)^2+1/(1+\mu)^3]$.
These calculations up to $\sigma^4$ suggest that similar infinite series in powers of $\mu$
exist for each power of $\sigma$, which can be resummed. I have however not been able to find a systematic
(diagrammatic ?) expansion to prove this to all orders in $\sigma$. With the considerations just given, one has
\begin{eqnarray}\label{avgfinss}
\overline N = \kappa [(1+\mu)^{-1} +(1+\mu)^{-2}\gamma \sigma^2+[(1+\mu)^{-2}+(1+\mu)^{-3}] \gamma^2 \sigma^4]  + 5 \gamma^3 \sigma^6 + {\cal O}(\sigma^8,S^{-1})\, , 
\end{eqnarray}
which matches Eq.~\eqref{avg1fin} when $\mu=0$. Care must be taken when taking into account the $\sigma^6$-term, because it is the zeroth-order term in an expansion in $\mu$. 
Including that term, Eq.~\eqref{avgfinss} is strictly valid only for $|\mu| \ll 1$.\\[-8mm]

\subsubsection{Second expansion}

The second Neumann expansion takes ${\mathbb X} = \mathbb{1+M}$ and ${\mathbb Y}= \delta {\mathbb A}$ in Eq.~\eqref{neumann}. I obtain, instead of Eqs.~\eqref{avg0} and \eqref{avgmu},
\begin{equation}\label{expandN2}
\overline N = \frac{\kappa}{1+\mu} \, \sum_{p=0}^\infty \sum_{j} (-1)^p \, \Big\langle \Big[ \delta \mathbb{A} \big({\mathbb 1}-\frac{{\mathbb M}}{1+\mu}\big)\Big]^p_{ij} \Big\rangle \, ,
\end{equation}
where I used $(\mathbb{1+M})^{-1}=(\mathbb{1}-\mathbb{M}/(1+\mu)$.
Taking into account terms up to $p=4$, I obtain
\begin{equation}\label{equi2mu}
\overline N = \kappa \, \left[(1+\mu)^{-1} +(1+\mu)^{-2}\gamma \sigma^2+[(1+\mu)^{-2}+(1+\mu)^{-3}] \gamma^2 \sigma^4 \right]+ {\cal O}(\sigma^6,S^{-1})\, .
\end{equation}
Remarkably,  I recover Eq.~\eqref{avgfinss} up to and including the $\sigma^4$ terms. Remember, however, that Eq.~\eqref{avgfinss} is derived 
under the assumption  that $|\mu|<1$ and $\sigma(1+|\gamma|)<1$, while Eq.~\eqref{equi2mu} is valid for an extended range, $\sigma (1+|\gamma|) \lesssim1$, i.e. with no restriction on $\mu$. 
This further suggests that a resummation of the perturbation series obtained 
with the first approach may give a more compact and elegant expression valid for any value of $\mu$, for $\sigma(1+|\gamma|)<1$. At this stage, this statement is however only a conjecture. 

\section{RMT calculation of the variance of the abundance} 

From Eq.~\eqref{equin},  the second moment of the abundance distribution is given by 
\begin{eqnarray}\label{sig0}
\langle (N_i^*)^2 \rangle_{\mu=0} &=&  \sum_{p_1,p_2} \sum_{j_1,j_2} (-1)^{p_1+p_2}  \Big\langle [\mathbb{A}^{p_1}]_{i j_1}   [\mathbb{A}^{p_2}]_{i j_2} k_{j_1} k_{j_2}  \Big\rangle 
= \sum_{p_1,p_2} \sum_{j_1,j_2} (-1)^{p_1+p_2}   \Big\langle [\mathbb{A}^{p_1}]_{i j_1}   [\mathbb{A}^{p_2}]_{i j_2} \Big\rangle \langle  k_{j_1} k_{j_2}  \rangle
\end{eqnarray}
\begin{wrapfigure}{r}{0.53\textwidth}
\begin{center}
\vspace{-1.4cm}
\includegraphics[width=0.46 \columnwidth]{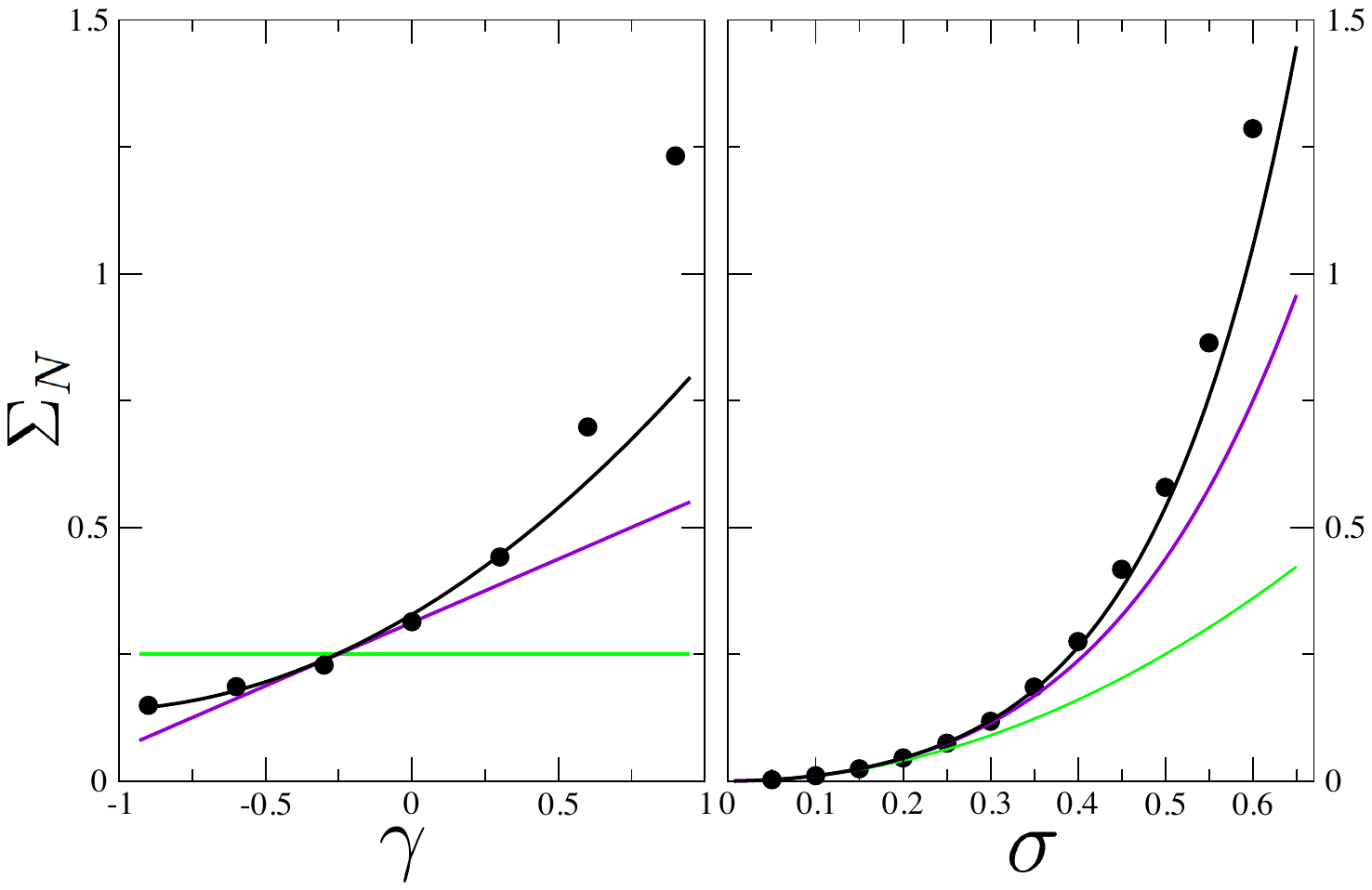}
\vspace{-1cm}
\caption{\label{fig:figS2}  Convergence of the series expansion of Eq.~\eqref{equivarkappa} for $\mu=0$ and $k_i \equiv \kappa =  1$ (hence $\chi=0$). 
The solid curves give Eq.~\eqref{equivarkappa}  truncated at the second (green), fourth (violet) and sixth (black) order in $\sigma$.\\[-10mm]}
 \end{center}
\end{wrapfigure}
where, as before, the average is taken over both a RMT distribution of interaction matrices as well as a distribution of carrying capacities, under 
 the assumption that the two are uncorrelated. 
 
The calculation is similar as for the average of the abundance, with similar sum rules as above restricting the number of Kroneckers giving contributions surviving the $S \rightarrow \infty$ limit.
The corresponding index contractions are similar to those discussed above, with however different sequences of indices in the terms of Eq.~\eqref{sig0} with $p_{1} \ne 0 \ne p_2$.
I calculated contributions up to order ${\cal O}(\sigma^6)$  for $\mu=0$, however I give here only details up to ${\cal O}(\sigma^4)$. 
I have (factors of two count multiplicities of terms, e.g. in Eq.~\eqref{sig0}, $p_1=2$, $p_2=0$ gives the same contribution as $p_1=0$, $p_2=2$),
\begin{eqnarray}\label{sig1}
\langle (N_i^*)^2 \rangle_{\mu=0} &=& \langle k_i^2 \rangle + \sum_{j_1,j_2} \Big\langle \mathbb{A}_{i j_1}   \mathbb{A}_{i j_2} \Big\rangle \langle k_{j_1} k_{j_2} \rangle
+ 2 \sum_{j,l}  \Big\langle \mathbb{A}_{il}   \mathbb{A}_{lj} \Big\rangle \langle k_{j} k_{i} \rangle
 + 2 \sum_{j,l_1,l_2,l_3}  \Big\langle \mathbb{A}_{i l_1}   \mathbb{A}_{l_1 l_2} \mathbb{A}_{l_2 l_3}   \mathbb{A}_{l_3 j} \Big\rangle \langle k_{j} k_{i} \rangle \nonumber \\
&&
+ 2 \sum_{j_1,j_2,l_1,l_2} \Big\langle \mathbb{A}_{i l_1}   \mathbb{A}_{l_1 l_2} \mathbb{A}_{l_2 j_1}   \mathbb{A}_{i j_2} \Big\rangle \langle k_{j_1} k_{j_2} \rangle 
+ \sum_{j_1,j_2,l_1,l_2} \Big\langle \mathbb{A}_{i l_1}   \mathbb{A}_{l_1 j_1} \mathbb{A}_{i l_2}   \mathbb{A}_{l_2 j_2} \Big\rangle\langle k_{j_1} k_{j_2} \rangle
+ {\cal O}(\sigma^6)  \nonumber \\
&=&  (\kappa^2+\chi^2) \, [1+\sigma^2+2 \gamma \sigma^2 + 4 \gamma^2 \sigma^4 + 4 \gamma \sigma^4 + (1+\gamma^2) \sigma^4 ]+ {\cal O}(\sigma^6,S^{-1}) \, ,
\end{eqnarray}
where in the square bracket in the last line, terms have been grouped to sequentially correspond to the six terms on the right hand side of the first equality.
The variance of the abundance for $\mu=0$, up to and including ${\cal O}(\sigma^6)$ is given by 
\begin{eqnarray}\label{equivarkappa}
\Sigma_N  &=&  \chi^2[1+2 \gamma \sigma^2+5 \gamma^2 \sigma^4+14 \gamma^3 \sigma^6] + (\chi^2+\kappa^2)  [\sigma^2+(1+4 \gamma) \sigma^4 + (1+6 \gamma + 14 \gamma^2) \sigma^6]   + {\cal O}(\sigma^8,S^{-1})\, . \qquad
\end{eqnarray}

The calculation of the variance of the abundance in the case $\mu \ne 0$ is tedious, though straightforward. 
Terms with no $\mu$-dependence are multiplied by $\langle k_i^2 \rangle = \kappa^2 + \chi^2$,
while those with a $\mu$-dependence are multiplied by $\langle k_i \rangle^2 = \kappa^2$. I obtain Eq.~({\color{red}5}) in the main text, using the expansion of Paragraph~\ref{firstexpan}.
Fig.~\ref{fig:figS2} illustrates the convergence of the series expansion in Eq.~\eqref{equivarkappa} to the numerically obtained results, as orders in $\sigma$ are added. 

\section{RMT calculation of higher moments} 

Skewness and kurtosis can be computed starting from the third and fourth moments of the abundance distribution,\\[-4mm]
\begin{subequations}
\begin{eqnarray} 
\langle (N_i^*)^3 \rangle &=&  \sum_{p_1,p_2,p_3} \sum_{j_1,j_2,j_3}  (-1)^{p_1+p_2+p_3} \Big\langle [\mathbb{A}^{p_1}]_{i j_1}   [\mathbb{A}^{p_2}]_{i j_2} [\mathbb{A}^{p_3}]_{i j_3} k_{j_1} k_{j_2}  k_{j_3} \Big\rangle \, , \\
\langle (N_i^*)^4 \rangle &=&  \sum_{p_1,p_2,p_3,p_4} \sum_{j_1,j_2,j_3,j_4} (-1)^{p_1+p_2+p_3+p_4 }    \Big\langle [\mathbb{A}^{p_1}]_{i j_1}   [\mathbb{A}^{p_2}]_{i j_2} [\mathbb{A}^{p_3}]_{i j_3} [\mathbb{A}^{p_3}]_{i j_4}  k_{j_1} k_{j_2}  k_{j_3} k_{j_4}  \rangle \, .
\end{eqnarray}
\end{subequations}
Calculating them up to and including ${\cal O}(\sigma^6)$ is not more complicated than for the average and variance, though it becomes intricate with $\mu \ne 0$  and $\chi \ne 0$.
To sketch the calculations, I therefore restrict myself to the case $\mu=0$ and $k_i \equiv \kappa =  1$, $\chi=0$, i.e. 
\begin{subequations}
\label{skewkurt}
\begin{eqnarray} 
\label{skew}
\langle (N_i^*)^3 \rangle &=&  \tilde{\sum_{p_1,p_2,p_3}} \sum_{j_1,j_2,j_3} \Big\langle [\mathbb{A}^{p_1}]_{i j_1}   [\mathbb{A}^{p_2}]_{i j_2} [\mathbb{A}^{p_3}]_{i j_3} \Big\rangle \, , \\
\label{kurt}
\langle (N_i^*)^4 \rangle &=&  \tilde{\sum_{p_1,p_2,p_3,p_4}} \sum_{j_1,j_2,j_3,j_4}     \Big\langle [\mathbb{A}^{p_1}]_{i j_1}   [\mathbb{A}^{p_2}]_{i j_2} [\mathbb{A}^{p_3}]_{i j_3} [\mathbb{A}^{p_4}]_{i j_4}  \Big \rangle \, ,
\end{eqnarray}
\end{subequations}
where $\tilde{\sum}$ indicates that the sum is restricted to $\sum_i p_i =2 p$, i.e. the sum of powers is an even number. 

For the third moment, and up to and including ${\cal O}(\sigma^6)$, we need to calculate terms with $(p_1,p_2,p_3)$ such that $p_1+p_2+p_3 \le 6$. 
As an example, the $(p_1,p_2,p_3)=(2,1,1)$ contribution to Eq.~\eqref{skew} reads, including its multiplicity of 3 [i.e. $(1,2,1)$ and $(1,1,2)$ give the same contribution],
\begin{eqnarray} 
3 \sum_{j_1,j_2,j_3} \sum_{l} \langle \mathbb{A}_{il} \mathbb{A}_{lj_1} \mathbb{A}_{ij_2} \mathbb{A}_{ij_3} \rangle = 3 \sum_{j_1,j_2,j_3} \sum_{l} \Big[  \langle \mathbb{A}_{il} \mathbb{A}_{lj_1} \rangle \langle  \mathbb{A}_{ij_2}  \mathbb{A}_{ij_3} \rangle +  \langle \mathbb{A}_{il} \mathbb{A}_{ij_2} \rangle \langle  \mathbb{A}_{lj_1}  \mathbb{A}_{ij_3} \rangle
+  \langle \mathbb{A}_{il} \mathbb{A}_{ij_3} \rangle \langle  \mathbb{A}_{lj_1}  \mathbb{A}_{ij_2} \rangle \Big] \, . 
\end{eqnarray}
Interactions give a prefactor $\sigma^4/S^2$ to all three contributions. There are four index sums, therefore, only contributions with two index contractions at most survive  the $S \rightarrow \infty$ limit.
The first contribution gives $\langle \mathbb{A}_{il} \mathbb{A}_{lj_1} \rangle \langle  \mathbb{A}_{ij_2}  \mathbb{A}_{ij_3} \rangle = \gamma \, (\sigma^4/S^2)  \delta_{ij_1} \delta_{j_2j_3}$
None of the last two  contributions  survives the $S \rightarrow \infty$ limit, because both have three index contractions. I obtain
\begin{eqnarray} 
3 \sum_{j_1,j_2,j_3} \sum_{l} \langle \mathbb{A}_{il} \mathbb{A}_{lj_1} \mathbb{A}_{ij_2} \mathbb{A}_{ij_3} \rangle = 3 \gamma \sigma^4.
\end{eqnarray}
Other terms are calculated in a similar way. As before, the ${\cal O}(\sigma^6)$ terms have been computed in parallel by a symbolic algebra code and the just described analytical calculation.
All terms contributing to the third moment, Eq.~\eqref{skew}, and fourth moment, Eq.~\eqref{kurt}, are listed in Tables~\ref{table3} and \ref{table4}, respectively. \\

\begin{table}[!h]
\begin{tabular}{|c|c|c|c|}
 \hline
Contributions to & $(p_1,p_2,p_3)$  & Multiplicity &  Contribution \\
 Eq.~\eqref{skew}  & & & \\
    \hline
&(0,0,0) & 1& 1 \\
& (1,1,0) & 3& $3 \sigma^2  $\\
 & (2,0,0) & 3& $3 \gamma \sigma^2  $\\
 & (2,1,1) & 3&  $3 \gamma \sigma^4  $\\
 & (2,2,0) & 3&  $3 (1+\gamma^2) \sigma^4  $\\
 & (3,1,0) & 6&  $12 \gamma \sigma^4  $\\
 & (4,0,0) & 3& $6 \gamma^2 \sigma^4  $\\
  \hline
\end{tabular}
\begin{tabular}{|c|c|c|}
 \hline
 $(p_1,p_2,p_3)$  & Multiplicity &  Contribution \\
& & \\
    \hline
(2,2,2) & 1& $(3 \gamma + \gamma^3) \sigma^6 $\\
 (3,2,1) & 6& $12 \gamma^2  \sigma^6  $\\
 (3,3,0) & 3& $3 (1+4 \gamma^2) \sigma^6  $\\
  (4,1,1) & 3&  $6 \gamma^2 \sigma^6  $\\
  (4,2,0) & 6&  $6 (3\gamma+2\gamma^3) \sigma^6  $\\
  (5,1,0) & 6&  $30 \gamma^2 \sigma^6  $\\
  (6,0,0) & 3& $15 \gamma^3 \sigma^6  $\\
  \hline
\end{tabular}
\caption{Contributions to the third moment, Eq.~\eqref{skew}, up to and including ${\cal O}(\sigma^6)$. All contributions listed in the fourth and seventh columns already include their multiplicity as listed in the 
third and sixth columns.}\label{table3}
\end{table}

\begin{table}[!h]
\begin{tabular}{|c|c|c|c|}
 \hline
Contributions to & $(p_1,p_2,p_3,p_4)$  & Multiplicity &  Contribution \\
 Eq.~\eqref{kurt}  & & & \\
  \hline
& (0,0,0,0) & 1& 1 \\
 & (1,1,0,0) &  6 & $6 \sigma^2  $\\
 & (2,0,0,0) & 4 & $4 \gamma \sigma^2  $\\
 & (1,1,1,1) & 1&  $ 3  \sigma^4  $\\
 & (2,1,1,0) & 12&  $ 12 \gamma \sigma^4  $\\
 & (2,2,0,0) & 6&  $6 (1+\gamma^2) \sigma^4  $\\
 & (3,1,0,0) & 12&  $24 \gamma \sigma^4  $\\
 & (4,0,0,0) & 4& $8 \gamma^2 \sigma^4  $\\
  &   & &  $  $\\
  \hline
\end{tabular}
\begin{tabular}{|c|c|c|}
 \hline
  $(p_1,p_2,p_3,p_4)$  &Multiplicity &  Contribution \\
 & &  \\
  \hline
 (2,2,1,1) & 6& $6 (1+\gamma^2) \sigma^6$  \\
  (2,2,2,0) &  4 & $4(3 \gamma + \gamma^3) \sigma^6  $\\
  (3,1,1,1) & 4&  $ 24  \gamma \sigma^6  $\\
  (3,2,1,0) & 24&  $48  \gamma^2 \sigma^6  $\\
    (3,3,0,0) & 6 & $6 (1+4  \gamma^2) \sigma^6  $\\
  (4,1,1,0) & 12&  $24 \gamma^2 \sigma^6  $\\
  (4,2,0,0) & 12 & $ 12 (3 \gamma + 2 \gamma^3) \sigma^6$ \\
  (5,1,0,0) & 12&  $60 \gamma^2 \sigma^6  $\\
  (6,0,0,0) & 4& $20 \gamma^3 \sigma^6  $\\
  \hline
\end{tabular}
\caption{Contributions to the fourth moment, Eq.~\eqref{kurt}, up to and including ${\cal O}(\sigma^6)$. All contributions listed in the fourth and seventh columns include their multiplicity as listed in the 
third and sixth columns.}\label{table4}
\end{table}

Skewness and kurtosis are obtained from third and fourth moments as 
\begin{eqnarray} 
\label{sk_defs}
{\cal S}=\langle (N_i^*-\overline N)^3\rangle/\Sigma_N^{3/2} \, , \:\:\:\:\:\:\:
{\cal K}=\langle (N_i^*-\overline N)^4\rangle/\Sigma_N^{2}  \, .
 \end{eqnarray}
 Using the contributions listed in Table~\ref{table3}, I obtain a vanishing skewness, ${\cal S}={\cal O}(\sigma^8)$.
 With that result, and the contributions listed in Table~\ref{table4}, I further obtain ${\cal K}=3 +{\cal O}(\sigma^8,S^{-1})$. Therefore, up to and including terms ${\cal O}(\sigma^6)$, skewness and kurtosis 
 indicate that we have a Gaussian distribution of species abundances. 

This argument is easily extended to the case with distributed carrying capacities, $k_i \not \equiv \kappa $, as well as to $\mu \ne 0$,
with the same conclusion, that the abundance distribution is Gaussian up to order ${\cal O}(\sigma^6)$. I finally note that, 
as for the kurtosis, when calculating even standardized moments of higher orders,
one obtains pairwise contractions already existing in lower moments, with combinatorial factors of $(2 p -1) !!$, relating the $2p^{\rm th}$ moment to the variance, that correspond to the Gaussian case~[{\color{blue} 51}].

\section{From the distribution of abundances to the probability of negative abundances} 

So far I have found that species abundances of fixed-point solutions of Eq.~({\color{red}3}) are distributed according to the Gaussian distribution,
\begin{equation}
P(N_i^*)=\exp[-(N_i^*-\overline N)^2/2 \Sigma_N]/\sqrt{2 \pi \Sigma_N} \, , 
\end{equation}
whose average and variance are given in the perturbation expansion of Eqs.~({\color{red}4}) and ({\color{red}5}). With $S$ species, one has, by normalization of probabilities,
\begin{eqnarray}
1 &\equiv&  \left[ \int_{-\infty}^0 P(N_i^*) {\rm d} N +  \int^{\infty}_0 P(N_i^*) {\rm d} N \right]^S  = \sum_{N_e=0}^S 
\left(\begin{array}{c}
S \\ N_e
\end{array} \right) \left[ \int_{-\infty}^0 P(N_i^*) {\rm d} N \right]^{N_e} \, \left[  \int^{\infty}_0 P(N_i^*) {\rm d} N \right]^{S-N_e}\,\, , 
\end{eqnarray}
Each term in the sum on the right-hand side gives the probability to have $N_e$ species with negative abundances. Noting that 
\begin{eqnarray}
\int_{-\infty}^0 P(N_i^*) {\rm d} N &=& \left(1-{\rm erf}[\overline N/\sqrt{2 \Sigma_N}]\right)\big/2 \, , \\
\int^{\infty}_0 P(N_i^*) {\rm d} N &=&\left(1+{\rm erf}[\overline N/\sqrt{2 \Sigma_N}]\right)\big/2 \, ,
\end{eqnarray}
I obtain Eq.~({\color{red}6}) in the main text. The  probability distribution of negative abundances depends on a single parameter only, $\eta=\overline N/\sqrt{2 \Sigma_N}$. Together with the numerical
observation that species with negative abundances are those that go extinct under the time-evolution of Eq.~({\color{red}1}), this observation leads to the conjecture of a single-parameter scaling 
of species extinctions. 

\section{Negative abundances at unfeasible fixed points vs. extinctions} 

When species go extinct, the dynamics of Eq.~({\color{red}1}) flows to a new fixed-point in a space of reduced dimensionality, where extinct species are removed because they
have $N_i(t \rightarrow \infty ) \rightarrow 0$.
The new fixed-point equation is
\begin{equation}\label{equired}
[\mathbb{1}+\mathbb{A}]_{\rm red}  {\vec N^*}_{\rm red} =  \vec{k}_{\rm red} \, ,
\end{equation}
where the subscript "red" indicates a dimensionality reduction following the removal of extinct species. 
On the other hand, identifying species that get extinct under the dynamics of Eq.~({\color{red}1}) with those that have negative abundances according to the fixed-point equation, is tantamount to considering
the fixed-point solution in the complete space, 
\begin{equation}\label{equicons}
[\mathbb{1}+\mathbb{A}]  {\vec N^*} =  \vec{k}\, ,
\end{equation}
and, somewhat arbitrarily, setting to zero all negative abundance components.
The two procedures, are different and mathematically, there must be a 
discrepancy in abundances, $N^*_{{\rm red},i} - N^*_i \ne 0$ for surviving species.
In this paragraph, I show that this shift  is on the order of $S_e/S$, with the number $S_e$ of extinct species. Therefore, as long as $S_e \ll S$, this shift is negligibly small and 
it is justified to identify species with negative abundances from Eq.~\eqref{equi2} with extinct species. 

Consider that the system has parameters and a random realization of the interaction matrix leading to an unfeasible solution of Eq.~\eqref{equi2}, with $N^*_i > 0$ for $i \in {\cal S}_{s}$ and $N^*_i \le 0$ for $i \in {\cal S}_e$, ${\cal S}_s \cup {\cal S}_e = {\cal S}$. I construct a reduced fixed-point solution of 
Eq.~\eqref{equired}  as $N_{{\rm red},i}(t) = N^*_i + \delta N_i(t)$, $i \in {\cal S}_s$. From Eq.~({\color{red}1}), the shift $\delta N_i(t)$ evolves dynamically as
\begin{eqnarray}
 \delta \dot N_i = [N_{i}^*+\delta N_i] [k_i - (N_i^*+\delta N_i) - \sum_{j \in {\cal S}_s} \mathbb{A}_{ij} (N_j^*+\delta N_j) ] \, , \;\;\;\;\;\;\; i \in {\cal S}_s\, .
\end{eqnarray}
Under the assumption that the shift does not lead to further extinctions -- the condition for this will be given below -- this
leads to a new fixed point defined by time-independent shifts $\delta N^*_i$ satisfying
\begin{eqnarray}\label{deltaN}
k_i - (N_i^*+\delta N_i^*) - \sum_{j \in {\cal S}_s} \mathbb{A}_{ij} (N_j^*+\delta N_j^*) = 0 \, .
\end{eqnarray}
Because $\vec{N}^*$ is a solution of the unrestricted fixed-point Eq.~\eqref{equi2}, Eq.~\eqref{deltaN} 
can be rewritten as
\begin{eqnarray}\label{shift1}
\delta N_i^* + \sum_{j \in {\cal S}_s} \mathbb{A}_{ij}  \delta N_j^* = \sum_{j \in {\cal S}_e} \mathbb{A}_{ij}  N^*_j  \, , \;\;\;\;\;\;\; i \in {\cal S}_s \, .
\end{eqnarray}
This gives a linear relation between the fixed-point shifts $\delta N_i^*$ of the surviving species and the negative abundances  at the unfeasible fixed-point. 
To evaluate the shifts, I note that the left-hand side of Eq.~\eqref{shift1} can be rewritten $[(\mathbb{1}+\mathbb{A})_{\rm red} \delta \vec{N}^*]_i$. The shifts are then 
obtained by inverting $[\mathbb{1}+\mathbb{A}]_{\rm red}$. Using a Neumann expansion and rewriting $N_j^*$ on the right-hand side of
Eq.~\eqref{shift1} with Eq.~\eqref{expandN}, I finally obtain
\begin{eqnarray}\label{shift2}
\delta N_i^* = \sum_{p,p'=0}^\infty \, \sum_{j \in {\cal S}_s} \sum_{k \in {\cal S}_e} 
\sum_{l=1}^S (-1)^{p+p'} \, [\mathbb{A}_{\rm red}^p]_{ij} \mathbb{A}_{jk} [\mathbb{A}^{p'}]_{kl} k_l \, , \;\;\;\;\;\;\; i \in {\cal S}_s \, ,
\end{eqnarray}
with the reduced interaction matrix $\mathbb{A}_{\rm red} \in \mathbb{R}^{S_s \times S_s}$ between the $S_s$ surviving species. 

The RMT-averaged shift is calculated from Eq.~\eqref{shift2} as usual. Only contributions
with odd $p+p'$ survive the RMT average. I calculate the six contributions up to and including $p+p'=3$. I obtain
\begin{eqnarray}\label{shift3}
\delta \overline N = -\kappa \frac{S_e}{S} \left\{ \gamma \sigma^2 + 4 \gamma^2 \sigma^4 \frac{S_s}{S} -2 \mu^2 -\mu^4 \left[1+\frac{S_s}{S} + \frac{S_s^2}{S^2} + \frac{S_s^3}{S^3} \right]\right\}  \, .
\end{eqnarray}
The shift is small, as long as $S_e/S \ll 1$. I conclude that the shifted fixed-point with dynamically-induced extinctions remains very close to the unfeasible fixed point projected onto the subspace of 
species with positive abundances, as long as the relative number of extinct species remains small. In this limit, these shifts are small and lead to no statistically significant additional extinctions.

\end{document}